\documentclass[twocolumn,showpacs,preprintnumbers,amsmath,amssymb]{revtex4}
\usepackage{graphicx}% Include figure files
\usepackage{dcolumn}% Align table columns on decimal point
\usepackage{bm}% bold math
\begin{document}

%\preprint{APS/123-QED}

\title{The suppression of superconductivity in Mn substituted MgCNi$_{3}$}

\author{A. Das }
\email{adas@apsara.barc.ernet.in}
 \altaffiliation[On leave from ]{Solid State Physics Division, Bhabha Atomic Research Centre, Mumbai 400 085, India}%Lines break automatically or can be forced with \\
% \email{adas@apsara.barc.ernet.in}
\author{R.K. Kremer}%
% \email{Second.Author@institution.edu}
\affiliation{ Max-Planck-Institut f\"{u}r
Festk\"{o}rperforschung,\ 70569 Stuttgart, Germany}
% \textbackslash\textbackslash }

%\author{Charlie Author}
% \homepage{http://www.Second.institution.edu/~Charlie.Author}
%\affiliation{
%Second institution and/or address\\
%This line break forced% with \\
%}%

\date{\today}% It is always \today, today,
             %  but any date may be explicitly specified

\begin{abstract}
We report the effect of doping Mn in the isostructural
MgCNi$_{3-x}$Mn$_{x}$ (x = 0-0.05) compounds. Magnetic
susceptibility, resistivity, magneto-resistance, and specific heat
studies show evidence of localized moments and Kondo effect in
samples with x$\neq$0. The rapid suppression of superconductivity
($\sim$~ -21K/at.\% Mn) in these compounds is a consequence of
 pair breaking effects due to moment formation on Mn.
\end{abstract}

% PACS, the Physics and Astronomy
                             % Classification Scheme.
%\keywords{Suggested keywords}%Use showkeys class option if keyword
                              %display desired
\maketitle

\section{\label{sec:level1}Introduction\protect\\}% The line
%break was forced {\lowercase{via} \textbackslash\textbackslash}%

The intermetallic compound MgCNi$_{3}$ has recently been reported
to exhibit superconductivity with T$_{C}$ $\sim$~8K \cite{1}. It
is related structurally to the borocarbides RT$_{2}$B$_{2}$C
(T=Ni, Pd) \cite{1a,1b} and may be viewed as three dimensional
analogue of these from the point of arrangement of the Ni atoms.
This structural relationship therefore, provides the possibility
of comparing the properties between these two. MgCNi$_{3}$ has
generated interest because of its simple perovskite structure and
absence of a ferromagnetic ground state inspite of a large
proportion of Ni per unit cell. The superconducting ground state
has raised the question of the role of Ni-3d bands and led to the
suggestion that MgCNi$_{3}$ to be a candidate for unconventional
superconductivity \cite{1c}. Electronic structure calculations
show Ni d-band derived density of states to dominate the
electronic states at the Fermi energy, E$_{F}$ \cite{2,3,4,5}. The
Mg and C atoms provide a contribution to the electronic density of
states (DOS) at about 1 eV below E$_{F}$. The proximity of the Ni
d band to E$_{F}$ may lead to spin fluctuations or magnetic order
which could eventually suppress superconductivity. The absence of
ferromagnetism in MgCNi$_{3}$ is shown to result from low values
of the Stoner factor, S = 0.43 \cite{3}, 0.64 \cite{4} and the
DOS. Similarly, absence of antiferromagnetism has been ruled out
due to lack of nesting features at the Fermi surface \cite{3}. The
electronic structure calculations also indicate that doping at one
of the three atomic sites could induce a large change in the
superconducting state. In particular, electronic structure
calculations predict that hole doping \cite{6} could drive the
system to a ferromagnetic ground state while electron doping
\cite{5} could lead to Fermi surface nesting and subsequently
induce a charge density wave or spin density wave instability.

The nature of superconductivity in MgCNi$_{3}$ is far from being
resolved, experimentally. MgCNi$_{3}$ crystallizes in the cubic
perovskite structure with C atoms in the body center position
surrounded by Ni atoms at the face center position, and Mg atoms
at cube corners. The superconducting phase as determined from
neutron diffraction experiments is found to be
MgC$_{0.96}$Ni$_{3}$ for T$_{C}$ $\sim$~8K \cite{1}. The
sensitivity of the superconducting properties to the Mg and C
content has been studied and it was found that T$_{C}$ decreases
systematically with decreasing C content \cite{7,8}.
Superconductivity is identified with the $\beta$-MgCNi$_{3}$ cubic
phase with cell constant $\sim$~3.816$\AA$ \cite{9}. The
superconductivity is described within the BCS model. However the
strength of electron-phonon coupling reported varies from weak
\cite{10}, moderate \cite{11} to strong coupling \cite{12,13}. NMR
studies indicate that the nature of pairing to be s-wave
\cite{13A}, and Hall effect results show that the carriers are of
electron type \cite{10}. However, tunnelling experiments indicate
that MgCNi$_{3}$ is a strong coupling superconductor with non
s-wave pairing symmetry \cite{12}.

There are a number of theoretical \cite{14,15,16} and experimental
\cite{17,18,19} studies on the effect of doping at the Ni site in
MgCNi$_{3}$. Doping the Ni site with other transition metals leads
to changes in the degree of filling of partially filled Ni d-bands
which influences the superconducting properties. Experimentally,
effect of doping in MgCNi$_{3-x}$T$_{x}$ (T = Cu, Co, Fe, and Mn)
is found to suppress superconductivity in varied ways.
Substituting Ni by Cu (limited to ~3\% by the solubility of Cu
into Ni ) leads to a decrease of T$_{C}$ from 7 to 6K \cite{17}.
Hayward et al. \cite{17} observe that doping with 1\% Co
suppresses bulk superconductivity. Resistivity measurements by
Kumary et al. \cite{18} and Ren et al. \cite{19} on Co doped
samples show that T$_{C}$ systematically decreases from 7.7K to
7.1K for x between 0 and 0.4. In the case of partially
substituting Ni with Fe, it is found that T$_{C}$ initially
increases \cite{18}. Magnetic ordering has not been reported in
either Fe or Co substituted samples, in strong contrast to
theoretical predictions. A brief report on Mn substitution
indicates that superconductivity is suppressed in samples with x =
0.125 and 0.5 \cite{19}. In this communication we present results
which demonstrate that doping by Mn leads to suppression of
superconductivity much more pronounced than previously observed.
We identify, this behavior to be related with the formation of
localized moments on Mn and a concomitant Kondo effect.

\section{\label{sec:level2}Experimental Details\protect\\}

In our study polycrystalline samples with nominal composition
Mg$_{1.2}$C$_{1.5}$Ni$_{3-x}$Mn$_{x}$ ( x = 0 - 0.05) were
investigated. Powder samples were prepared by solid state reaction
following the method reported by Ren et al. \cite{9} for x = 0.
The starting materials used were powders of Mg ( 99.8\%, Alfa
Aesar), Ni (99.99\%, Aldrich) glassy Carbon (Alfa Aesar) and Mn (
99.99\%, Alfa Aesar). Stoichiometric amounts of these materials
were mixed together, pressed into a pellet and sealed in a Ta tube
under high purity Argon. Excess of Mg was used to compensate for
the volatility of Mg and excess of C was found necessary to arrive
at samples with highest T$_{C}$. The mixtures were slowly heated
to 600$^{\circ}$C and kept for 2 hours and heated to
950$^{\circ}$C and again kept for 2 hours. After cooling, the
pellets were reground and repeatedly annealed at 950$^{\circ}$C
for 5 hours. X-ray patterns were recorded with Mo radiation
($\lambda$ = 0.70930 $\AA$) on a Stoe powder diffractometer. DC
magnetization measurements were carried out in a MPMS SQUID
magnetometer (Quantum Design). AC resistivity measurements
($\nu$=19 Hz) were done in a PPMS (Quantum Design) on in-situ
pressed pellets using a home built sapphire cell with Pt contacts
employing the van der Pauw method. Specific heat measurements were
carried out in the PPMS on powdered samples mixed with Apiezon
grease. Addenda from the platform and the grease as determined
from separate runs were subtracted.

\section{\label{sec:level3}Results and Discussion\protect\\}

Figure 1 shows typical x-ray diffraction patterns of some of our
samples. All the samples are isostructural. Negligibly small
impurity reflections from MgO were observed (2$\theta$ $\approx$
19$^{\circ}$). The presence of a minute amount of unreacted Ni is
inferred from the observation of a small kink at $\sim$~625K in
the high-field susceptibility data, coinciding with the Curie
temperature of Ni. The lattice parameter of the parent compound
(x=0) is 3.8110(2)\ensuremath{\AA}. Comparing this value with the
reported variation of T$_{C}$ versus carbon concentration
\cite{7}, we estimate that the carbon content in this sample is
$\sim$~0.97. The slightly lower carbon concentration results in a
reduction of T$_{C}$ from the optimal value. With addition of Mn
(x$\neq$0) the cell constant increases as shown in the inset of
Fig.1. The increase in the cell parameters with addition of Mn is
consistent with the larger atomic radii of Mn as compared to Ni.
We note that in case of Co and Fe doping, very little or no change
in cell parameters had been observed \cite{18}.

\begin{figure}
\resizebox{0.50\textwidth}{!}{
\includegraphics{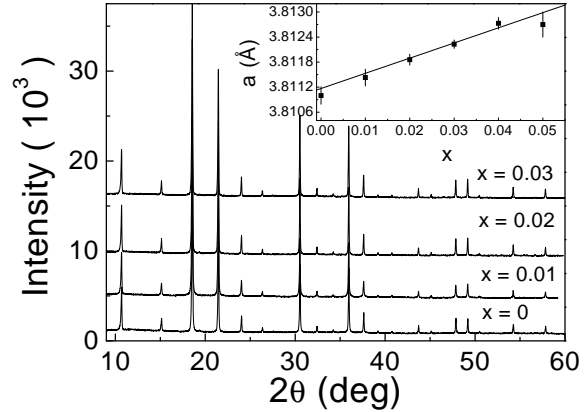}% Here is how to import EPS art
}
%\vspace{5 cm}
\caption{\label{fig:epsart} X-ray diffraction patterns ($\lambda$
= 0.70930$\AA$) of MgCNi$_{3-x}$Mn$_{x}$ (x=0, 0.01, 0.02, 0.03).
The patterns are shifted vertically for clarity. The inset shows
the variation of the lattice constant a with concentration.}
\end{figure}

\begin{figure}[b]
\resizebox{0.50\textwidth}{!}{
\includegraphics{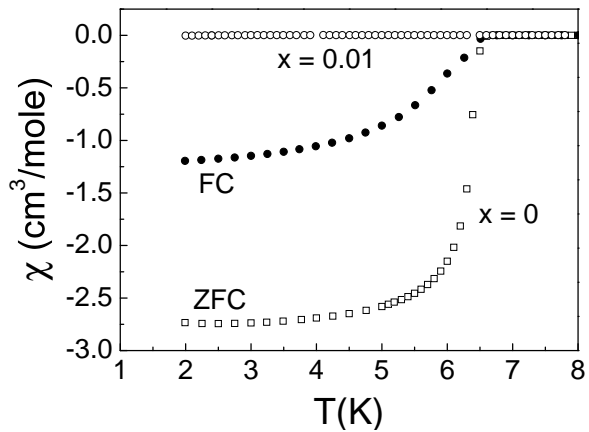}% Here is how to import EPS art
}
%\vspace{5 cm}
\caption{\label{fig:epsart} Magnetic susceptibility $\chi$ as a
function of temperature measured in magnetic field, H=6Oe for
samples MgCNi$_{3-x}$Mn$_{x}$ (x=0, 0.01). Zero-field-cooled (ZFC)
and field-cooled (FC) branches are marked.}
\end{figure}

The variation of the susceptibility as a function of temperature
for x=0 and 0.01 is shown in figure 2. The parent compound (x=0)
shows a T$_{C}$$^{onset}$ of $\approx$6.6K. It exhibits a fairly
sharp transition and below 4K reaches approximately 100$\%$ of the
superconducting volume fraction. On substituting Ni with Mn, even
for the lowest concentration x = 0.01 (0.3 at.$\%$ Mn),
superconductivity is completely suppressed. A very weak
diamagnetic signal was observed at 1.8K which correlates with
resistivity data discussed later. For higher Mn concentration no
evidence of superconductivity was found. Additionally, ac
susceptibility measurements were carried out down to 0.3K which
gave no evidence of superconducting and/or magnetic ordering in Mn
doped samples. This rapid suppression of both T$_{C}$
($\sim$~-21K/at.$\%$Mn) and volume fraction, on addition of Mn is
very different from that reported in the case of Co, Fe, and Cu
substitutions, where only a marginal decrease in T$_{C}$ had been
observed. Substitution of Mn in YNi$_{2}$B$_{2}$C also leads to
only a marginal decrease of 0.7K/at.$\%$ of Mn \cite{20}. The
suppression of superconductivity in Mn doped MgCNi$_{3}$ can be
attributed to the formation of local moments the evidence of which
comes from paramagnetic susceptibility and resistivity
measurements described in the following.

The paramagnetic susceptibility, $\chi$ as a function of
temperature measured in a field of 1kOe is shown in figure 3. For
x = 0 the variation is nearly temperature independent. The
magnitude of $\chi$ $\sim$~3.5$\times$10$^{-4}$ cm$^{3}$/mole is
close to that reported by Hayward et al. \cite{17}. For samples
with x$>$0 a large increase in $\chi$ at low temperatures is
observed. The temperature dependence of $\chi$ follows a
Curie-Weiss like behavior and has been fitted to a relation of the
form $\chi = \chi_{0}+C/(T-\theta)$ in the temperature range 5 -
300K. $\chi_{0}$ $\sim$ ~4$\times10^{-4}$ cm$^{3}$/mole represents
the sum of all temperature independent contributions. The value of
$\theta$ are $\sim$~4K. From the  Curie constant, C the
paramagnetic moment is estimated. The effective magnetic moment
per Mn obtained ($\sim$~6$\mu_{B}$) is close to the expected free
ion value 5.9$\mu_{B}$ for Mn$^{2+}$ (S = 5/2). The slight
discrepancy results from the uncertainty in the final composition.
This is the first report of observation of Curie-Weiss behavior
and therefore of a local moment formation in these samples. The
field dependence of magnetization at 5K for samples x= 0.03, 0.04,
and 0.05 is shown in Figure 4. There is no hysteresis observed for
any composition or temperature. The variation of magnetization is
non linear but does not exhibit saturation and does not approach
the free ion value expected for S = 5/2. For comparison the
expected behavior for S = 5/2 is also plotted. Magnetization as a
function of magnetic field measured at 2, 5, and 10K is shown in
inset of figure 4 for a typical sample, x = 0.05. The data at
different temperatures do not collapse into a single curve
indicating that the doped Mn moments are magnetically interacting.

\begin{figure}[t]
\resizebox{0.50\textwidth}{!}{
\includegraphics{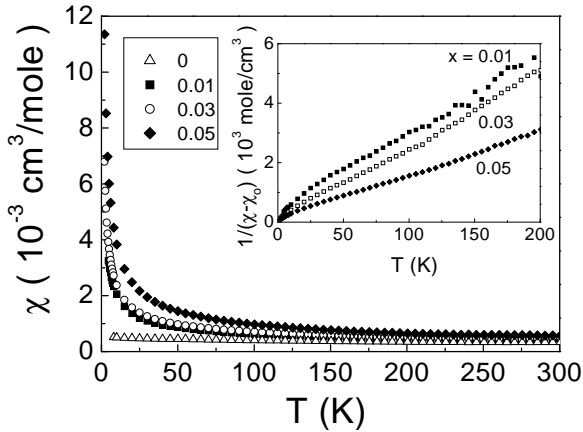}% Here is how to import EPS art
}
%\vspace{5 cm}
\caption{\label{fig:epsart} The normal state magnetic
susceptibility $\chi$ as a function of temperature measured in
magnetic field, H=1000Oe for samples MgCNi$_{3-x}$Mn$_{x}$ (x=0,
0.01, 0.03, 0.05). The inset shows temperature variation of
inverse susceptibility for x = 0.01, 0.03, and 0.05.}
\end{figure}

\begin{figure}[t]
\resizebox{0.50\textwidth}{!}{
\includegraphics{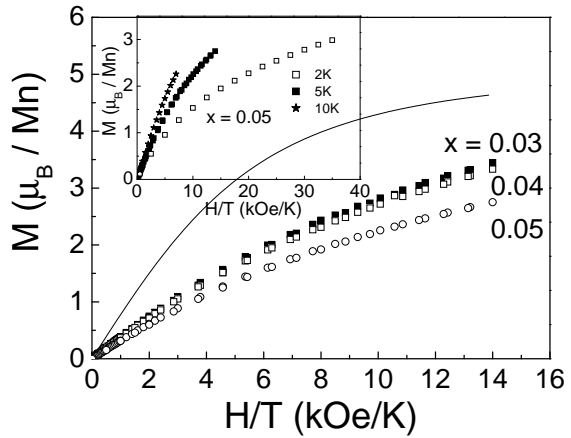}% Here is how to import EPS art
}
%\vspace{5 cm}
\caption{\label{fig:epsart} Magnetization as a function of field
of MgCNi$_{3-x}$Mn$_{x}$ (x = 0.03, 0.04, 0.05) at 5K. The
continuous line is the expected variation from a Brillioun
function for spin S=5/2. The inset shows the variation of
magnetization with field at temperatures 2, 5, and 10K for
x=0.05.}
\end{figure}

\begin{figure}[b]
\resizebox{0.50\textwidth}{!}{
\includegraphics{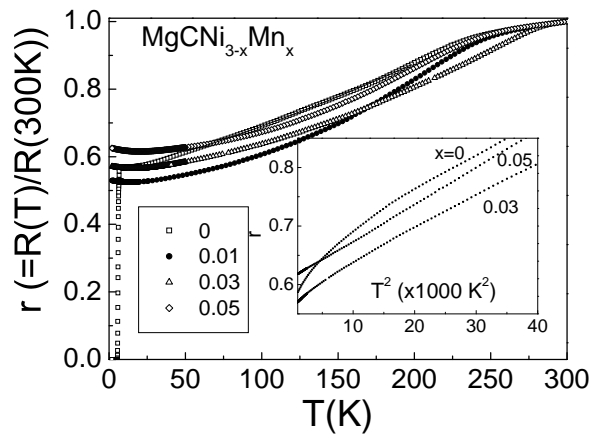}% Here is how to import EPS art
}
%\vspace{5 cm}
\caption{\label{fig:epsart} The variation of normalized resistance
r (=R(T)/R(300K)) with temperature for samples
MgCNi$_{3-x}$Mn$_{x}$ (x = 0, 0.01, 0.03, 0.05). the continuous
line through the data points of x = 0 is the fit to equation 1
(described in the text). The inset shows the T$^{2}$ dependence of
r followed in the case of x = 0, 0.03 and 0.05.}
\end{figure}

Indications for a Kondo behavior and associated spin fluctuation
effects are observed in the temperature dependence of the
resistivity. Figure 5 shows the plot of normalized resistance r (=
R(T)/R(300K)) as a function of temperature. The resistance
decreases with temperature and exhibit a metallic behavior. While
the undoped compound shows superconductivity below $\sim$~6.8K,
the Mn containing compounds exhibit a minimum in r(T) at
T$_{min}$, below which r increases again with lowering of T. The
residual resistivity ratio (RRR) $\sim$~1.8 for x = 0 is about the
same as reported by Waelte et al. \cite{17} but lower than that
reported by Kumary et al. \cite{18} The low value of RRR along
with low value of T$_{C}$ and cell parameters correlate with the
deficiency of carbon. The absolute value of the resistivity in our
samples are between $\sim$~10-20 m$\Omega$ cm which are about an
order of magnitude larger than found by others \cite{17,18}. We
attribute this difference to grain boundary resistance in the non
sintered pellets used in the experiment. The temperature
dependence of r between 50 and 200K could be described by the
relation
\begin{equation}
r(T) = A+BT^{2}+C\left(T \over \theta \right)^{5}\
\int_{0}^{\frac{\theta}{T}}dx\frac{x^{5}}{[(e^{x}-1)(1-e^{-x})]}
\end{equation}
where A is a temperature independent term, the second term arises
from spin fluctuation effects and the third is the Bloch
Gr\"{u}neissen function. It gives a reasonably good fit to the
data. For x = 0, the value of $\theta_{D}$ $\sim$~238K extracted
from the fit is slightly lower than that obtained from specific
heat measurements $\theta_{D}$ $\sim$~280K. In the Mn containing
samples an increase in the coefficient of the T$^{2}$ term from
2.6$\times10^{-6}$ for x=0 to $\sim$~6.5$\times10^{-6}$ (K$^{-2}$)
for x=0.05 is observed. The T$^{2}$ dependence is shown in the
inset of Fig.5. The enhancement of the T$^{2}$ term in x$\neq$0 is
attributed to an increase in scattering from spin fluctuations
\cite{21}.

\begin{figure}[b]
\resizebox{0.50\textwidth}{!}{
\includegraphics{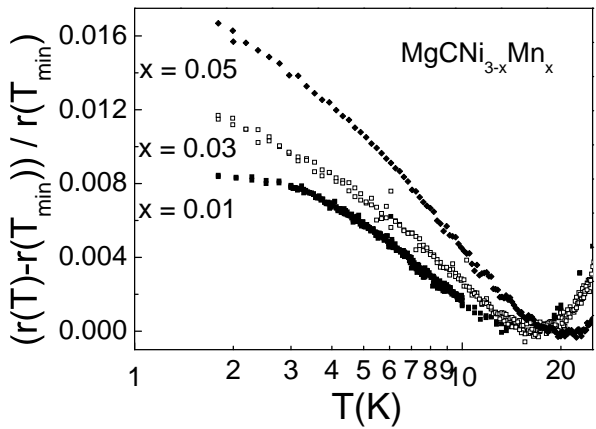}% Here is how to import EPS art
}
%\vspace{5 cm}
\caption{\label{fig:epsart} The variation of
(r(T)-r(T$_{min}$))/r(T$_{min}$) with temperature for samples
MgCNi$_{3-x}$Mn$_{x}$ (x = 0.01, 0.03, 0.05) in the region below
T$_{min}$.}
\end{figure}

For samples with x$\neq$0 the position of the resistivity minimum,
T$_{min}$ increases from $\sim$~16K for x =0.01 to $\sim$~20K for
x=0.05, as shown in Figure 6. And similarly, the depth of the
minimum (r(1.8K)-r(T$_{min}$))/r(T$_{min}$) increases from
$\sim$~0.8$\%$ for x=0.01 to $\sim$~1.7$\%$ for x=0.05. The
temperature dependence of r(T) below T$_{min}$ follows a -$\ln$T
behavior in all the cases. The slope 1/c dr/d$\ln$T decreases from
9$\times10^{-3}$ for x= 0.01 to 2.7$\times10^{-3}$ (K
at.\%)$^{-1}$ for x=0.05. All these features are consistent and
indicate a dominating role of Kondo effect \cite{22,23} in the low
temperature electronic properties of these samples. The variation
of the slope with concentration suggests that the impurities are
not isolated but magnetically interacting which supports the
conclusion drawn from M(H) curves. Evidence for a Kondo effect is
also found from magnetoresistance (MR) measurements displayed in
Figure 7. The MR, $\Delta$R/R(0) (=(R(H)-R(0))/R(0)) as a function
of field is negative and follows a $\sim$~H$^{2}$ dependence. The
negative MR results from partial alignment of impurity spins which
reduces the spin flip scattering. A correlation between the MR and
magnetization in a Kondo system has been shown to exist and
described by the expression \cite{24,25}
\begin{equation}
\Delta R = R(H,T)-R(0,T) =
{\frac{3\pi}{2E_{F}}}\frac{m}{e^{2}\hbar}cVJ^{2}M^{2}
\end{equation}
where m and e are the charge and mass of the electron, V is the
atomic volume, J the s-d exchange constant, c the atomic
concentration of impurities, E$_{F}$ the Fermi energy, and M is
the magnetization in $\mu_{B}$/atom. In the inset of Fig. 7 we
show such a correlation to exist in these samples. Using v$_{F}$ =
1.5$\times$10$^{7}$ cm$^{2}$/s \cite{13} in the above expression
we obtain the value for $\mid$ J $\mid$ $\sim$~0.1 eV.  This value
of J is comparable to those reported in the case of dilute alloys
of Mn in Cu \cite{25} and Mn in PdSb \cite{25A}.

\begin{figure}[t]
\resizebox{0.50\textwidth}{!}{
\includegraphics{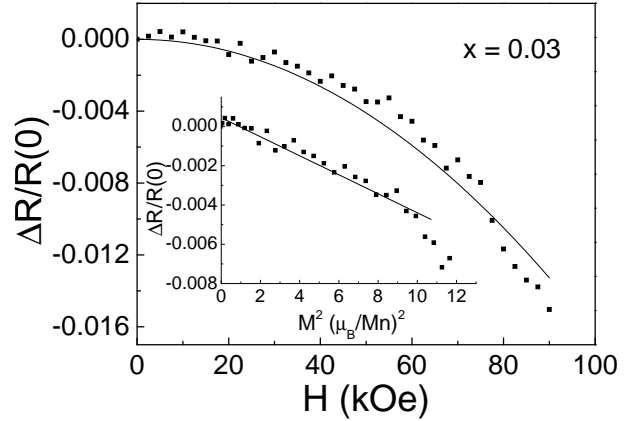}% Here is how to import EPS art
}
%\vspace{5 cm}
\caption{\label{fig:epsart} The variation of magneto-resistance,
$\Delta$R/R(0) with field at T=2K for MgCNi$_{2.97}$Mn$_{0.03}$.
The continuous line shows the $\sim$~H$^{2}$ dependence. The inset
shows the variation of $\Delta$R/R(0) with M$^{2}$.}
\end{figure}

Specific heat measurements were carried out in the temperature
range 1.8 to 20K. The data could be fitted (for x = 0, in the
range T $>$T$_{C}$) to an expression of the form
\begin{equation}
C_{P}(T)=\gamma T + C_{L} + C_{E} + C_{Sch}
\end{equation}
where $\gamma$ is the Sommerfeld constant, C$_{L}$, C$_{E}$, and
C$_{Sch}$ are the Debye, Einstein, and a Schottky term for
magnetic contribution, respectively. The Einstein term takes into
account the 13 meV octahedral rotational mode \cite{4}. The fitted
parameters are shown in Table 1. The variation of C$_{P}$ with
temperature for x = 0 is shown in inset of Fig.8. The
superconducting anomaly at $\sim$~7K indicates bulk
superconductivity and is in agreement with preceeding
measurements. However, the anomaly in comparison to those reported
in the literature \cite{13,15} is somewhat broader possibly
indicating a distribution of T$_{C}$ due to defects. The value of
$\gamma$ $\sim$~10.7 (mJ/mole K per Ni atom) obtained from the fit
of the zero field data is in good agreement with other reports in
the literature.  The absence of superconductivity in x$\neq$0 is
also evident from specific heat measurements confirming the bulk
nature of the effect. The temperature variation of C$_{P}$ is
shown in Fig.8 for a typical sample x=0.03. At low temperatures
the increase observed in C$_{P}$/T versus T$^{2}$ plot arises from
the Schottky anomaly. The maximum is estimated to be at 0.4K and
the energy splitting is $\sim$~1K. The values of $\gamma$ appear
to increase (beyond the uncertainty of fitting) with increasing Mn
concentration. This observation indicates conclusively that
reduction in T$_{C}$ with addition of Mn is due to the pair
breaking interaction which is of magnetic origin and not
electronic. Similar studies of specific heat in carbon deficient
samples show that the reduction in T$_{C}$ correlate with the
decrease in $\gamma$ \cite{8}. On application of magnetic field,
the maximum in the Schottky anomaly shifts to higher temperatures.
Inset (b) in Fig.8 shows the difference plot, C$_{P}$(90
kOe)-C$_{P}$(0 kOe) versus T for a sample with x = 0.03. A broad
maximum is observed at $\sim$~9K. Qualitatively, the nature of the
curve could be reproduced by assuming a Schottky anomaly for a  2
level system. The excess specific heat on application of field is
understood to arise from the removal of the degeneracy of the
Kondo quasi bound state. The entropy obtained is within 25\% of
the expected value (R ln2) for S=1/2.

\begin{figure}
\resizebox{0.5\textwidth}{!}{
\includegraphics{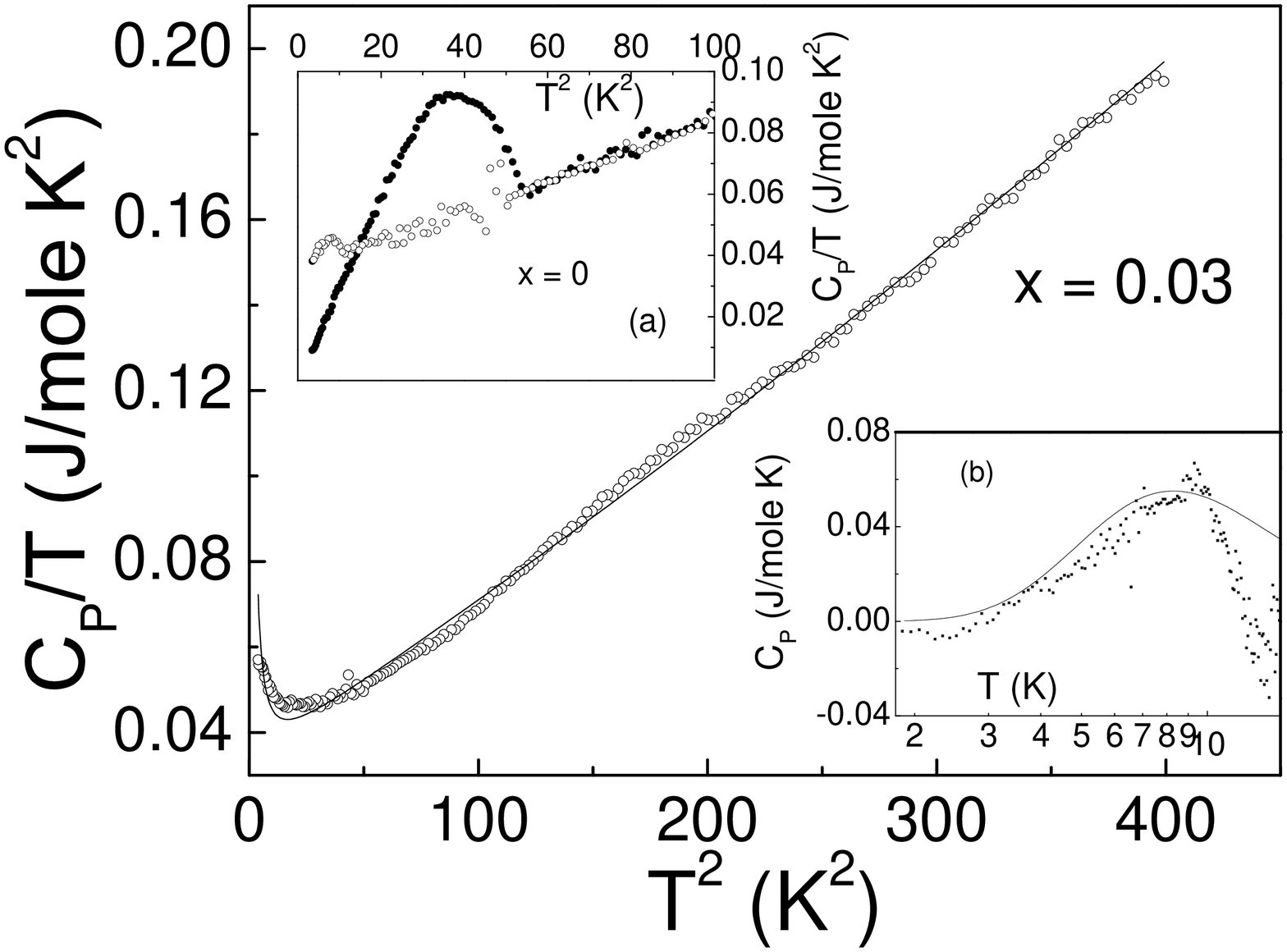}% Here is how to import EPS art
}
%\vspace{1 cm}
 \caption{\label{fig:epsart} The plot of C$_{P}$/T versus T$^{2}$ for
MgCNi$_{2.97}$Mn$_{0.03}$. The continuous line is the fit to
equation 3. In the inset (a) is shown the variation of C$_{P}$/T
versus T$^{2}$ for MgCNi$_{3}$ ($\bullet$H=0, $\circ$H=90kOe). In
the inset (b) is shown the difference, C$_{P}$(H=90
kOe)-C$_{P}$(H=0) against T for MgCNi$_{2.97}$Mn$_{0.03}$. The
solid curve is generated assuming a 2 level Schottky effect.}
\end{figure}

\begin{table}
\caption{\label{tab:table1} Parameters Debye temperature
($\theta$$_{D}$), Sommerfeld parameter ($\gamma$), Einstein
temperature ($\theta$$_{E}$) and Schottky energy
($\delta$$_{Schottky}$) obtained from fitting C$_{P}$(T) data to
equation 3 for samples MgCNi$_{3-x}$Mn$_{x}$ ( x = 0, 0.01, 0.03,
0.05) }
\begin{ruledtabular}
\begin{tabular}{ccccc}
x &$\theta$$_{D}$ (K)&$\gamma$ (mJ/mole
K$^{2}$)&$\theta$$_{E}$ (K)& $\delta$$_{Schottky}$ (K) \\
\hline
0.0 & 256 & 32.0& 149 & - \\
0.01 & 288 & 28.2 & 150 & 0.8\\
0.03 & 294 & 32.4 & 153 & 1.0\\
0.05 & 296 & 35.0 & 164 & 1.6\\
\end{tabular}
\end{ruledtabular}
\end{table}

From susceptibility measurements the Kondo temperature, T$_{K}$
was estimated using the expression,
$\chi(T)=g^{2}\mu_{B}^{2}S(S+1)/3.66(k_{B}T+4.5T_{K})$ \cite{22}
valid for T$>$T$_{K}$. Comparing the $\theta$ values we find
T$_{K}\sim~1K$ in these compositions. It is of interest to note
that we also find that T$_{C}$ is only marginally reduced to 5.5K
in the case of MgCNi$_{2.95}$V$_{0.05}$ and
MgCNi$_{2.95}$Cr$_{0.05}$. In addition, no Curie-Weiss behavior is
observed. The absence of moment in the case of Cr and V doped
samples is contrary to the results arrived at by electronic
structure calculations by Granada et al. \cite{16}, where it is
shown that moments are formed more easily as the difference in
atomic number, $\Delta$Z, between Ni and transition metal
increases. Therefore, we conclude local moments form only in the
case of Mn and they greatly inhibit superconductivity. The
reduction of T$_{C}$ in the presence of paramagnetic impurities
has been worked out by Abrikosov and Gorkov (AG) \cite{26}. Within
the AG theory the variation of T$_{C}$ is given by the relation
\begin{equation}
\ln\frac{T_{C}}{T_{C0}}=\psi(\frac{1}{2})-\psi(\frac{1}{2}+\rho)
\end{equation}
where T$_{C0}$=T$_{C}$ for x=0, $\psi$ is the digamma function,
and $\rho$ (= $\alpha$/2$\pi$k$_{B}$T$_{C}$) contains $\alpha$,
the pair breaking energy. In terms of J and S, $\alpha$ is
expressed as $\alpha$ = 4$\pi$cN(E$_{F}$)J$^{2}S(S+1)$.
M\"{u}ller-Hartmann and Zittartz (MZ) \cite{27} in their theory on
the effect of Kondo behavior on superconductivity have shown that
the variation of T$_{C}$ with concentration is markedly different
from the AG theory. In this case the pair breaking strength
depends upon T$_{C}$  and $\rho$ is expressed as
\begin{equation}
\rho
=\overline{c}\frac{T_{C0}}{T_{C}}\frac{\pi^{2}S(S+1)}{ln^{2}T_{C}/T_{K}+\pi^{2}S(S+1)}
\end{equation}

As a result, the reduction of T$_{C}$ with impurity concentration
is a function of T$_{K}$/T$_{C}$. For the case,
T$_{K}$$\ll$T$_{C}$ a characteristic reentrant
superconducting-normal behavior is observed \cite{27A}. In Figure
9 we plot the variation of T$_{C}$ with normalized concentration,
$\overline{c}$ =c/(2$\pi$)$^{2}$N(E$_{F}$)T$_{C}$ for the case of
S = 5/2 using the estimated values of J and T$_{K}$ for
MgCNi$_{3}$ in the expressions from AG and MZ theory. In the
presence of Kondo effect the suppression of T$_{C}$ is much more
rapid than in the case of normal paramagnetic impurities. The
critical concentration as observed in the figure $\overline{c}$ =
0.15 ($\simeq$0.35 at.\%) is in good agreement with our
observation of absence of superconductivity in samples with 0.3
at.\% Mn. Electronic structure calculations for MgCNi$_{3}$
estimate the moment on doped Co atoms to be
$\sim~10^{-5}\mu_{B}$/atom and attribute the suppression of
superconductivity to spin fluctuations and strong d-d covalent
interactions. Similar calculations have been reported for the case
of Mn substitution with x=0.042 and show that Mn forms a magnetic
moment of 1.06$\mu_{B}$ \cite{15}. As against Co, the rapid
suppression of T$_{C}$ in Mn is ascribed to the overlap of the Mn
d states with the majority band of Ni states. The systematic
decrease in the value of R(300K)/R(T$_{min}$) with increase in Mn
concentration indicates that the virtual bound states which form
with Mn doping are close to E$_{F}$. Such enhanced pair breaking
effects on introduction of Mn in constrast to other transition
metals have been observed earlier in case of TM-Al alloys
\cite{28}. We believe  that Coulomb repulsion, occupation number
of d electrons and Kondo effect play a role in suppressing
superconductivity on substituting Ni with other transition metals
in MgCNi$_{3}$.

\begin{figure}[t]
\resizebox{0.50\textwidth}{!}{
\includegraphics{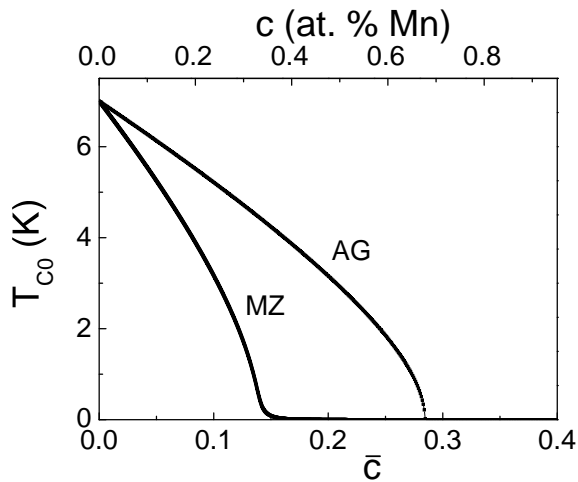}% Here is how to import EPS art
}
%\vspace{5 cm}
\caption{\label{fig:epsart} The plot of T$_{C}$ versus normalized
concentration $\overline{c}$ (=c/(2$\pi$)$^{2}$N(E$_{F})$T$_{C}$)
for the case of S = 5/2 using the estimated values of J and
T$_{K}$ in the expressions from AG and MZ theory.}
\end{figure}

\section{\label{sec:level1}Conclusion\protect\\}

In summary, we show that partially substituting Ni with Mn
dramatically alters superconductivity in MgCNi$_{3}$ as compared
to substitution of Ni with other transition metals. Doping by as
little as 0.3 at.\% of Mn completely suppresses superconductivity.
The Mn containing samples exhibit Curie-Weiss behavior in
temperature dependence of susceptibility and Kondo effect which
leads to rapid suppression of superconductivity.

\begin{acknowledgments}
We thank Eva Bruecher and Gisela Siegle for their assistance in
carrying out the various measurements. One of us (AD) acknowledges
Max-Planck Society fellowship for financial support.
\end{acknowledgments}

\bibliography{text}

\begin{thebibliography}{34}
\expandafter\ifx\csname natexlab\endcsname\relax\def\natexlab#1{#1}\fi
\expandafter\ifx\csname bibnamefont\endcsname\relax
  \def\bibnamefont#1{#1}\fi
\expandafter\ifx\csname bibfnamefont\endcsname\relax
  \def\bibfnamefont#1{#1}\fi
\expandafter\ifx\csname citenamefont\endcsname\relax
  \def\citenamefont#1{#1}\fi
\expandafter\ifx\csname url\endcsname\relax
  \def\url#1{\texttt{#1}}\fi
\expandafter\ifx\csname urlprefix\endcsname\relax\def\urlprefix{URL }\fi
\providecommand{\bibinfo}[2]{#2}
\providecommand{\eprint}[2][]{\url{#2}}

\bibitem[{\citenamefont{{T. He} et~al.}(2001)\citenamefont{{T. He}, {Q. Huang},
  {A.P. Ramirez}, {Y. Wang}, {K.A. Regan}, {N. Rogado}, {M.A. Hayward}, {M.K.
  Haas}, {J.J. Slusky}, {K. Inumara} et~al.}}]{1}
\bibinfo{author}{\bibnamefont{{T. He}}}, \bibinfo{author}{\bibnamefont{{Q.
  Huang}}}, \bibinfo{author}{\bibnamefont{{A.P. Ramirez}}},
  \bibinfo{author}{\bibnamefont{{Y. Wang}}},
  \bibinfo{author}{\bibnamefont{{K.A. Regan}}},
  \bibinfo{author}{\bibnamefont{{N. Rogado}}},
  \bibinfo{author}{\bibnamefont{{M.A. Hayward}}},
  \bibinfo{author}{\bibnamefont{{M.K. Haas}}},
  \bibinfo{author}{\bibnamefont{{J.J. Slusky}}},
  \bibinfo{author}{\bibnamefont{{K. Inumara}}}, \bibnamefont{et~al.},
  \bibinfo{journal}{Nature} \textbf{\bibinfo{volume}{411}}, \bibinfo{pages}{54}
  (\bibinfo{year}{2001}).

\bibitem[{\citenamefont{{R. Nagarajan} et~al.}(1994)\citenamefont{{R.
  Nagarajan}, {C. Mazumdar}, {Z. Hossain}, {S.K. Dhar}, {K.V. Gopalakrishnan},
  {L.C. Gupta}, {C. Godart}, {P.D. Padalia}, and {R. Vijaraghavan}}}]{1a}
\bibinfo{author}{\bibnamefont{{R. Nagarajan}}},
  \bibinfo{author}{\bibnamefont{{C. Mazumdar}}},
  \bibinfo{author}{\bibnamefont{{Z. Hossain}}},
  \bibinfo{author}{\bibnamefont{{S.K. Dhar}}},
  \bibinfo{author}{\bibnamefont{{K.V. Gopalakrishnan}}},
  \bibinfo{author}{\bibnamefont{{L.C. Gupta}}},
  \bibinfo{author}{\bibnamefont{{C. Godart}}},
  \bibinfo{author}{\bibnamefont{{P.D. Padalia}}}, \bibnamefont{and}
  \bibinfo{author}{\bibnamefont{{R. Vijaraghavan}}}, \bibinfo{journal}{Phys.
  Rev. Lett.} \textbf{\bibinfo{volume}{72}}, \bibinfo{pages}{274}
  (\bibinfo{year}{1994}).

\bibitem[{\citenamefont{{R.J. Cava} et~al.}(1994)\citenamefont{{R.J. Cava}, {H.
  Takagi}, {H.W. Zandbergen}, {J.J. Krajewski}, {W.F. Peck Jr.}, {T. Siegrist},
  {B. Batlogg}, {R.B. van Dover}, {R.J. Felder}, {K. Mizuhashi} et~al.}}]{1b}
\bibinfo{author}{\bibnamefont{{R.J. Cava}}}, \bibinfo{author}{\bibnamefont{{H.
  Takagi}}}, \bibinfo{author}{\bibnamefont{{H.W. Zandbergen}}},
  \bibinfo{author}{\bibnamefont{{J.J. Krajewski}}},
  \bibinfo{author}{\bibnamefont{{W.F. Peck Jr.}}},
  \bibinfo{author}{\bibnamefont{{T. Siegrist}}},
  \bibinfo{author}{\bibnamefont{{B. Batlogg}}},
  \bibinfo{author}{\bibnamefont{{R.B. van Dover}}},
  \bibinfo{author}{\bibnamefont{{R.J. Felder}}},
  \bibinfo{author}{\bibnamefont{{K. Mizuhashi}}}, \bibnamefont{et~al.},
  \bibinfo{journal}{Nature} \textbf{\bibinfo{volume}{367}},
  \bibinfo{pages}{252} (\bibinfo{year}{1994}).

\bibitem[{\citenamefont{Prozorov et~al.}()\citenamefont{Prozorov, Snezhko, He,
  and {R.J. Cava}}}]{1c}
\bibinfo{author}{\bibfnamefont{H.}~\bibnamefont{Prozorov}},
  \bibinfo{author}{\bibfnamefont{A.}~\bibnamefont{Snezhko}},
  \bibinfo{author}{\bibfnamefont{T.}~\bibnamefont{He}}, \bibnamefont{and}
  \bibinfo{author}{\bibnamefont{{R.J. Cava}}}, \eprint{Cond.
  Matt./0302431(v1)}.

\bibitem[{\citenamefont{Szajek}(2001)}]{2}
\bibinfo{author}{\bibfnamefont{A.}~\bibnamefont{Szajek}}, \bibinfo{journal}{J.
  Phys.:Condens. Matter} \textbf{\bibinfo{volume}{13}}, \bibinfo{pages}{L595}
  (\bibinfo{year}{2001}).

\bibitem[{\citenamefont{{S.B. Dugdale} and {T. Jarlborg}}(2001)}]{3}
\bibinfo{author}{\bibnamefont{{S.B. Dugdale}}} \bibnamefont{and}
  \bibinfo{author}{\bibnamefont{{T. Jarlborg}}}, \bibinfo{journal}{Phys.\ Rev.
  B} \textbf{\bibinfo{volume}{64}}, \bibinfo{pages}{100508}
  (\bibinfo{year}{2001}).

\bibitem[{\citenamefont{Singh and Mazin}(2001)}]{4}
\bibinfo{author}{\bibfnamefont{D.}~\bibnamefont{Singh}} \bibnamefont{and}
  \bibinfo{author}{\bibfnamefont{I.}~\bibnamefont{Mazin}},
  \bibinfo{journal}{Phys.\ Rev. B} \textbf{\bibinfo{volume}{64}},
  \bibinfo{pages}{140507} (\bibinfo{year}{2001}).

\bibitem[{\citenamefont{{J.H. Shim and S.K. Kwon and R.I. Min}}(2001)}]{5}
\bibinfo{author}{\bibnamefont{{J.H. Shim and S.K. Kwon and R.I. Min}}},
  \bibinfo{journal}{Phys.\ Rev. B} \textbf{\bibinfo{volume}{64}},
  \bibinfo{pages}{180510} (\bibinfo{year}{2001}).

\bibitem[{\citenamefont{{H. Rosner and R. Weht, M.D. Johannes and W.E. Pickett
  and E. Tossatti}}(2002)}]{6}
\bibinfo{author}{\bibnamefont{{H. Rosner and R. Weht, M.D. Johannes and W.E.
  Pickett and E. Tossatti}}}, \bibinfo{journal}{Phys.\ Rev. Lett.}
  \textbf{\bibinfo{volume}{88}}, \bibinfo{pages}{027001}
  (\bibinfo{year}{2002}).

\bibitem[{\citenamefont{{T.G. Amos and Q. Huang, J.W. Lynn and T. He and R.J.
  Cava}}(2002)}]{7}
\bibinfo{author}{\bibnamefont{{T.G. Amos and Q. Huang, J.W. Lynn and T. He and
  R.J. Cava}}}, \bibinfo{journal}{Solid State Commun.}
  \textbf{\bibinfo{volume}{121}}, \bibinfo{pages}{73} (\bibinfo{year}{2002}).

\bibitem[{\citenamefont{{L. Shan} et~al.}()\citenamefont{{L. Shan}, {K. Xia},
  {Z.Y. Lin}, {H.H. Wen}, {Z.A. Ren}, {G.C. Che}, and {Z.X. Zhao}}}]{8}
\bibinfo{author}{\bibnamefont{{L. Shan}}}, \bibinfo{author}{\bibnamefont{{K.
  Xia}}}, \bibinfo{author}{\bibnamefont{{Z.Y. Lin}}},
  \bibinfo{author}{\bibnamefont{{H.H. Wen}}},
  \bibinfo{author}{\bibnamefont{{Z.A. Ren}}},
  \bibinfo{author}{\bibnamefont{{G.C. Che}}}, \bibnamefont{and}
  \bibinfo{author}{\bibnamefont{{Z.X. Zhao}}}, \eprint{Cond. Matt./0302116}.

\bibitem[{\citenamefont{{Z.A. Ren} et~al.}(2002)\citenamefont{{Z.A. Ren}, {G.C.
  Che}, {S.L. Jia}, {H. Chen}, {Y.M. Ni}, {G.D. Lin}, and {Z.X. Zhao}}}]{9}
\bibinfo{author}{\bibnamefont{{Z.A. Ren}}}, \bibinfo{author}{\bibnamefont{{G.C.
  Che}}}, \bibinfo{author}{\bibnamefont{{S.L. Jia}}},
  \bibinfo{author}{\bibnamefont{{H. Chen}}},
  \bibinfo{author}{\bibnamefont{{Y.M. Ni}}},
  \bibinfo{author}{\bibnamefont{{G.D. Lin}}}, \bibnamefont{and}
  \bibinfo{author}{\bibnamefont{{Z.X. Zhao}}}, \bibinfo{journal}{Physica C}
  \textbf{\bibinfo{volume}{371}}, \bibinfo{pages}{1} (\bibinfo{year}{2002}).

\bibitem[{\citenamefont{{S.Y. Li} et~al.}(2001)\citenamefont{{S.Y. Li}, {R.
  Fan}, {X.H. Chen}, {C.H. Wang}, {W.Q. Mo}, {K.Q. Ruan}, {Y.M. Xiang}, {X.G.
  Luo}, {H.T. Zhang}, {L. Li} et~al.}}]{10}
\bibinfo{author}{\bibnamefont{{S.Y. Li}}}, \bibinfo{author}{\bibnamefont{{R.
  Fan}}}, \bibinfo{author}{\bibnamefont{{X.H. Chen}}},
  \bibinfo{author}{\bibnamefont{{C.H. Wang}}},
  \bibinfo{author}{\bibnamefont{{W.Q. Mo}}},
  \bibinfo{author}{\bibnamefont{{K.Q. Ruan}}},
  \bibinfo{author}{\bibnamefont{{Y.M. Xiang}}},
  \bibinfo{author}{\bibnamefont{{X.G. Luo}}},
  \bibinfo{author}{\bibnamefont{{H.T. Zhang}}},
  \bibinfo{author}{\bibnamefont{{L. Li}}}, \bibnamefont{et~al.},
  \bibinfo{journal}{Phys.\ Rev.B} \textbf{\bibinfo{volume}{64}},
  \bibinfo{pages}{132505} (\bibinfo{year}{2001}).

\bibitem[{\citenamefont{{J.-Y. Lin} et~al.}()\citenamefont{{J.-Y. Lin}, {P.L.
  Ho}, {H.L. Huang}, {P.H. Lin}, {Y.-L. Zhang}, {R.-C. Yu}, {C.-Q. Jin}, and
  {H.D. Yang}}}]{11}
\bibinfo{author}{\bibnamefont{{J.-Y. Lin}}},
  \bibinfo{author}{\bibnamefont{{P.L. Ho}}},
  \bibinfo{author}{\bibnamefont{{H.L. Huang}}},
  \bibinfo{author}{\bibnamefont{{P.H. Lin}}},
  \bibinfo{author}{\bibnamefont{{Y.-L. Zhang}}},
  \bibinfo{author}{\bibnamefont{{R.-C. Yu}}},
  \bibinfo{author}{\bibnamefont{{C.-Q. Jin}}}, \bibnamefont{and}
  \bibinfo{author}{\bibnamefont{{H.D. Yang}}}, \eprint{Cond. Matt./0202034}.

\bibitem[{\citenamefont{{Z.Q. Mao} et~al.}()\citenamefont{{Z.Q. Mao}, {M.M.
  Rosario}, {K.D. Nelson}, {K. Wu}, {I.G. Deac}, {P. Schiffer}, and {Y.
  Liu}}}]{12}
\bibinfo{author}{\bibnamefont{{Z.Q. Mao}}}, \bibinfo{author}{\bibnamefont{{M.M.
  Rosario}}}, \bibinfo{author}{\bibnamefont{{K.D. Nelson}}},
  \bibinfo{author}{\bibnamefont{{K. Wu}}}, \bibinfo{author}{\bibnamefont{{I.G.
  Deac}}}, \bibinfo{author}{\bibnamefont{{P. Schiffer}}}, \bibnamefont{and}
  \bibinfo{author}{\bibnamefont{{Y. Liu}}}, \eprint{Cond. Matt./0105280(v3)}.

\bibitem[{\citenamefont{{A.W\"{a}lte} et~al.}()\citenamefont{{A.W\"{a}lte}, {H.
  Rosner}, {M.D. Johannes}, {G. Fuchs}, {K.-H. M\"{u}ller}, {A. Handstein}, {K.
  Nenkov}, {V.N. Narozhnyi}, {S.-L. Drechsler}, {S. Shulga} et~al.}}]{13}
\bibinfo{author}{\bibnamefont{{A.W\"{a}lte}}},
  \bibinfo{author}{\bibnamefont{{H. Rosner}}},
  \bibinfo{author}{\bibnamefont{{M.D. Johannes}}},
  \bibinfo{author}{\bibnamefont{{G. Fuchs}}},
  \bibinfo{author}{\bibnamefont{{K.-H. M\"{u}ller}}},
  \bibinfo{author}{\bibnamefont{{A. Handstein}}},
  \bibinfo{author}{\bibnamefont{{K. Nenkov}}},
  \bibinfo{author}{\bibnamefont{{V.N. Narozhnyi}}},
  \bibinfo{author}{\bibnamefont{{S.-L. Drechsler}}},
  \bibinfo{author}{\bibnamefont{{S. Shulga}}}, \bibnamefont{et~al.},
  \eprint{Cond. Matt./0208364}.

\bibitem[{\citenamefont{Singer}(2001)}]{13A}
\bibinfo{author}{\bibfnamefont{P.}~\bibnamefont{Singer}},
  \bibinfo{journal}{Phys.\ Rev. Lett.} \textbf{\bibinfo{volume}{87}},
  \bibinfo{pages}{257601} (\bibinfo{year}{2001}).

\bibitem[{\citenamefont{{In Gee Kim, Jae Il Lee, A.J. Freeman}}(2002)}]{14}
\bibinfo{author}{\bibnamefont{{In Gee Kim, Jae Il Lee, A.J. Freeman}}},
  \bibinfo{journal}{Phys.\ Rev. B} \textbf{\bibinfo{volume}{65}},
  \bibinfo{pages}{064525} (\bibinfo{year}{2002}).

\bibitem[{\citenamefont{{J.L. Wang, Y. Xu, Z. Zeng, Q.Q. Zheng, H.Q.
  Lin}}(2002)}]{15}
\bibinfo{author}{\bibnamefont{{J.L. Wang, Y. Xu, Z. Zeng, Q.Q. Zheng, H.Q.
  Lin}}}, \bibinfo{journal}{J. Appl. Phys.} \textbf{\bibinfo{volume}{91}},
  \bibinfo{pages}{8504} (\bibinfo{year}{2002}).

\bibitem[{\citenamefont{{C.M. Granada, C.M. de Silva, A.A. Gomes}}(2002)}]{16}
\bibinfo{author}{\bibnamefont{{C.M. Granada, C.M. de Silva, A.A. Gomes}}},
  \bibinfo{journal}{Sol. State Commun.} \textbf{\bibinfo{volume}{122}},
  \bibinfo{pages}{269} (\bibinfo{year}{2002}).

\bibitem[{\citenamefont{{M.A. Hayward} et~al.}(2001)\citenamefont{{M.A.
  Hayward}, {M.K. Haas}, {A.P. Ramirez}, {T. He}, {K.A. Regan}, {N. Rogado},
  {K. Inumaru}, and {R.J. Cava}}}]{17}
\bibinfo{author}{\bibnamefont{{M.A. Hayward}}},
  \bibinfo{author}{\bibnamefont{{M.K. Haas}}},
  \bibinfo{author}{\bibnamefont{{A.P. Ramirez}}},
  \bibinfo{author}{\bibnamefont{{T. He}}}, \bibinfo{author}{\bibnamefont{{K.A.
  Regan}}}, \bibinfo{author}{\bibnamefont{{N. Rogado}}},
  \bibinfo{author}{\bibnamefont{{K. Inumaru}}}, \bibnamefont{and}
  \bibinfo{author}{\bibnamefont{{R.J. Cava}}}, \bibinfo{journal}{Sol. State
  Commun.} \textbf{\bibinfo{volume}{119}}, \bibinfo{pages}{491}
  (\bibinfo{year}{2001}).

\bibitem[{\citenamefont{{T.G. Kumary} et~al.}(2002)\citenamefont{{T.G. Kumary},
  {J. Janaki}, {Awadesh Mani}, {S. Mathi Jaya}, {V.S. Sastry}, {Y. Hariharan},
  {T.S. Radhakrishnan}, and {M.C. Valsakumar}}}]{18}
\bibinfo{author}{\bibnamefont{{T.G. Kumary}}},
  \bibinfo{author}{\bibnamefont{{J. Janaki}}},
  \bibinfo{author}{\bibnamefont{{Awadesh Mani}}},
  \bibinfo{author}{\bibnamefont{{S. Mathi Jaya}}},
  \bibinfo{author}{\bibnamefont{{V.S. Sastry}}},
  \bibinfo{author}{\bibnamefont{{Y. Hariharan}}},
  \bibinfo{author}{\bibnamefont{{T.S. Radhakrishnan}}}, \bibnamefont{and}
  \bibinfo{author}{\bibnamefont{{M.C. Valsakumar}}}, \bibinfo{journal}{Phys.\
  Rev. B} \textbf{\bibinfo{volume}{65}}, \bibinfo{pages}{214518}
  (\bibinfo{year}{2002}).

\bibitem[{\citenamefont{{Z.A. Ren and G.C. Che and S.L. Jia and H. Chen and
  Y.M. Ni and Z.X. Zhao}}()}]{19}
\bibinfo{author}{\bibnamefont{{Z.A. Ren and G.C. Che and S.L. Jia and H. Chen
  and Y.M. Ni and Z.X. Zhao}}}, \eprint{Cond. Matt./0105366}.

\bibitem[{\citenamefont{{F.S. da Rocha and G.L.F. Fraga. D.E. Brandao and A.A.
  Gomes}}(2001)}]{20}
\bibinfo{author}{\bibnamefont{{F.S. da Rocha and G.L.F. Fraga. D.E. Brandao and
  A.A. Gomes}}}, \bibinfo{journal}{Physica C} \textbf{\bibinfo{volume}{363}},
  \bibinfo{pages}{41} (\bibinfo{year}{2001}).

\bibitem[{\citenamefont{Dugdale}(1977)}]{21}
\bibinfo{author}{\bibfnamefont{J.}~\bibnamefont{Dugdale}},
  \emph{\bibinfo{title}{The Electrical properties of Metals and Alloys}}
  (\bibinfo{publisher}{Edward Arnold}, \bibinfo{year}{1977}).

\bibitem[{\citenamefont{Heeger}(1969)}]{22}
\bibinfo{author}{\bibfnamefont{A.}~\bibnamefont{Heeger}},
  \emph{\bibinfo{title}{Solid State Physics}} (\bibinfo{publisher}{Academic
  Press}, \bibinfo{year}{1969}), vol.~\bibinfo{volume}{23}, p.
  \bibinfo{pages}{283}.

\bibitem[{\citenamefont{Daybell}(1973)}]{23}
\bibinfo{author}{\bibfnamefont{M.~D.} \bibnamefont{Daybell}},
  \emph{\bibinfo{title}{Magnetism}} (\bibinfo{publisher}{Academic Press},
  \bibinfo{year}{1973}), vol.~\bibinfo{volume}{V}, p. \bibinfo{pages}{121}.

\bibitem[{\citenamefont{M.-T.B\'{e}al-Monod and Weiner}(1968)}]{24}
\bibinfo{author}{\bibnamefont{M.-T.B\'{e}al-Monod}} \bibnamefont{and}
  \bibinfo{author}{\bibfnamefont{R.}~\bibnamefont{Weiner}},
  \bibinfo{journal}{Phys.\ Rev.} \textbf{\bibinfo{volume}{552}},
  \bibinfo{pages}{170} (\bibinfo{year}{1968}).

\bibitem[{\citenamefont{Monod}(1967)}]{25}
\bibinfo{author}{\bibfnamefont{P.}~\bibnamefont{Monod}},
  \bibinfo{journal}{Phys.\ Rev. Lett.} \textbf{\bibinfo{volume}{19}},
  \bibinfo{pages}{1113} (\bibinfo{year}{1967}).

\bibitem[{\citenamefont{{T.H. Geballe, B.T. Matthias, B. Caroli, E. Corenzwit,
  J.P. Maita and G.W. Hull}}(1968)}]{25A}
\bibinfo{author}{\bibnamefont{{T.H. Geballe, B.T. Matthias, B. Caroli, E.
  Corenzwit, J.P. Maita and G.W. Hull}}}, \bibinfo{journal}{Phys.\ Rev.}
  \textbf{\bibinfo{volume}{169}}, \bibinfo{pages}{457} (\bibinfo{year}{1968}).

\bibitem[{\citenamefont{Abrikosov and Gor'kov}(1961)}]{26}
\bibinfo{author}{\bibfnamefont{A.}~\bibnamefont{Abrikosov}} \bibnamefont{and}
  \bibinfo{author}{\bibfnamefont{L.}~\bibnamefont{Gor'kov}},
  \bibinfo{journal}{Soviet Phys. JETP} \textbf{\bibinfo{volume}{12}},
  \bibinfo{pages}{1243} (\bibinfo{year}{1961}).

\bibitem[{\citenamefont{{E. M\"{u}ller-Hartmann and J. Zittartz}}(1971)}]{27}
\bibinfo{author}{\bibnamefont{{E. M\"{u}ller-Hartmann and J. Zittartz}}},
  \bibinfo{journal}{Phys.\ Rev. Lett.} \textbf{\bibinfo{volume}{26}},
  \bibinfo{pages}{428} (\bibinfo{year}{1971}).

\bibitem[{\citenamefont{{W.A. Fertig} et~al.}(1972)\citenamefont{{W.A. Fertig},
  {A.C. Mota}, {L.E. DeLong}, {D. Wohlleben}, and {R. Fitzgerald}}}]{27A}
\bibinfo{author}{\bibfnamefont{M.}~\bibnamefont{{W.A. Fertig}}},
  \bibinfo{author}{\bibnamefont{{A.C. Mota}}},
  \bibinfo{author}{\bibnamefont{{L.E. DeLong}}},
  \bibinfo{author}{\bibnamefont{{D. Wohlleben}}}, \bibnamefont{and}
  \bibinfo{author}{\bibnamefont{{R. Fitzgerald}}}, \bibinfo{journal}{Sol. State
  Commun.} \textbf{\bibinfo{volume}{11}}, \bibinfo{pages}{829}
  (\bibinfo{year}{1972}).

\bibitem[{\citenamefont{S.M.Vonsovsky}(1982)}]{28}
\bibinfo{author}{\bibfnamefont{E.~K.} \bibnamefont{S.M.Vonsovsky},
  \bibfnamefont{Yu. A.~Izynmov}}, \emph{\bibinfo{title}{Superconductivity of
  Transition Metals}}, vol.~\bibinfo{volume}{27} of
  \emph{\bibinfo{series}{Springer Series in Solid-State Sciences}}
  (\bibinfo{publisher}{Springer-Verlag}, \bibinfo{year}{1982}).

\end{thebibliography}

%\pagebreak
\clearpage

\end{document}